\documentclass[twocolumn]{aastex62}

\usepackage{amsmath}
\usepackage{lipsum}
\usepackage{float}
\usepackage{afterpage}
\usepackage{makecell}
\usepackage{upgreek}
\usepackage{bm}
\usepackage{ulem}
\usepackage{lineno}

\newcommand{\noop}[1]{}

\graphicspath{{./}{figures/}}
\shorttitle{Circumbinary Disk Decoupling}
\shortauthors{Dittmann, Ryan, \& Miller}

\begin{document}

\title{The Decoupling of Binaries from Their Circumbinary Disks}

\correspondingauthor{Alexander J. Dittmann}
\email{dittmann@astro.umd.edu}

\author[0000-0001-6157-6722]{Alexander J.~Dittmann}
\affil{Department of Astronomy and Joint Space-Science Institute, University of Maryland, College Park, MD 20742-2421, USA}
\author[0000-0001-9068-7157]{Geoffrey Ryan}
\affil{Perimeter Institute for Theoretical Physics, 31 Caroline St. N., Waterloo, ON, N2L 2Y5, Canada}

\author[0000-0002-2666-728X]{M. Coleman Miller}
\affil{Department of Astronomy and Joint Space-Science Institute, University of Maryland, College Park, MD 20742-2421, USA}

\begin{abstract} 
We have investigated, both analytically and numerically, accreting supermassive black hole binaries as they inspiral due to gravitational radiation to elucidate the decoupling of binaries from their disks and inform future multi-messenger observations of these systems. 
Our numerical studies evolve equal-mass binaries from initial separations of $100 GM/c^2$ until merger, resolving scales as small as $\sim0.04 GM/c^2$, where $M$ is the total binary mass. 
Our simulations accurately capture the point at which the orbital evolution of each binary decouples from that of their circumbinary disk, and precisely resolve the flow of gas throughout the inspiral. 
We demonstrate analytically and numerically that timescale-based predictions overestimate the binary separations at which decoupling occurs by factors of $\sim3$, and illustrate the utility of a velocity-based decoupling criterion.
High-viscosity ($\nu\gtrsim0.03 GM/c$) circumbinary systems decouple late ($a_b\lesssim 15 GM/c^2$) and have qualitatively similar morphologies near merger to circumbinary systems with constant binary separations. Lower-viscosity circumbinary disks decouple earlier and exhibit qualitatively different accretion flows, which 
lead to precipitously decreasing accretion onto the binary. If detected, such a decrease may unambiguously identify the host galaxy of an ongoing event within a LISA error volume.
We illustrate how accretion amplitude and variability evolve as binaries gradually decouple from their circumbinary disks, and where decoupling occurs over the course of binary inspirals in the LISA band.  We show that, even when dynamically negligible, gas may leave a detectable imprint on the phase of gravitational waves. 
\end{abstract}

\section{Introduction} 
Nearly all massive galaxies are thought to harbor supermassive black holes (SMBHs) in their centers \citep[e.g.,][]{2013ARA&A..51..511K}, resulting in the formation of SMBH binaries after the mergers of such galaxies and subsequent dynamical relaxation \citep[e.g.,][]{1980Natur.287..307B,2013CQGra..30x4008M}. The central regions of post-merger galaxies likely abound with dense gas \citep[e.g.][]{1992ARA&A..30..705B,2005ApJ...620L..79S}, which may form accretion disks around the binaries therein. Interactions between the binary and circumbinary disk may rapidly shrink the binary orbit \citep[e.g.,][]{2009ApJ...700.1952H,2022MNRAS.513.6158D}, which is of particular importance when the binary is too tightly bound to inspiral efficiently due to gravitational interactions with field stars \citep{2003ApJ...596..860M,2003AIPC..686..201M} but too widely separated to inspiral efficiently due to gravitational wave (GW) emission. 

The evolution of accreting SMBH binaries is typically thought to span three phases: first, at early times and large binary separations, energy and angular momentum loss due to GW emission is negligible \citep{1964PhRv..136.1224P}. During this stage, the orbital evolution of the binary is governed by interactions with the circumbinary disk \citep[see, e.g.,][and references therein]{2022arXiv221100028L}. Binary torques open a cavity in the circumbinary disk \citep[e.g.,][]{1991ApJ...381..259L,2015MNRAS.452.2396M}, the size of which is $\sim1.75-3.25$ times larger than the binary semi-major axis depending on the binary mass ratio and disk viscosity \citep{2020MNRAS.499.3362R,2022MNRAS.513.6158D}. On short timescales, depending on the angular momentum profile of the accreting gas, accretion onto the binary may be either suppressed \citep{1994ApJ...421..651A,2016MNRAS.460.1243R} or enhanced, although this effect vanishes as the disk approaches a viscous steady state \citep{2017MNRAS.466.1170M,2022MNRAS.513.6158D}. During this first stage, the rate at which the binary inspirals is necessarily limited by that at which the disk viscously spreads.

Eventually, the rate of change in the binary semi-major axis due to GW emission becomes sufficiently great that the binary and the disk \emph{decouple}, and the orbital evolution of the binary is dominated by GWs. It has often been suggested that this stage occurs when the timescale for GW-driven inspiral ($\tau_{GW}$) becomes shorter than the viscous timescale ($\tau_\nu$) at the disk edge, and that this might lead to a dearth of gas during the SMBH merger, a cessation of emission at shorter wavelengths, and the interruption of jet activity \citep[e.g.,][]{2003MNRAS.340..411L,2005ApJ...622L..93M,2012PhRvL.109v1102F}.\footnote{Although often cited when invoking a decoupling criterion of $\tau_{GW}\approx\tau_\nu$, \citet{2002ApJ...567L...9A} actually suggested a velocity-based decoupling criterion.} 
We more generally define decoupling as the epoch of disk-binary evolution during which the cavity size evolves in time more slowly than the binary semi-major axis. Regardless, eventually the inspiral proceeds quickly enough due to GW emission that the circumbinary disk is effectively frozen in place. Immediately after the binary merges, the mass lost from and kick imparted to the central black hole in may result in prompt electromagnetic counterparts \citep[e.g.][]{2008ApJ...684..835S,2009ApJ...700..859O,2010MNRAS.404..947C}, and subsequently as the circumbinary disk spreads the system will rebrighten as a single active galactic nucleus \citep[e.g.][]{2005ApJ...622L..93M,2010PhRvD..81b4019S}. 

Previous two-dimensional simulations have followed binaries through inspiral (e.g. \citet{2015MNRAS.447L..80F} and \citet{2018MNRAS.476.2249T}), making pioneering predictions for variability during inspiral, but both used a single viscosity and assumed that the systems decouple when $\tau_{\rm GW}=\tau_\nu$. \footnote{Specifically, \citet{2015MNRAS.447L..80F} implicitly rescaled the system parameters so that $\tau_{\rm GW}=\tau_\nu$ for their initial conditions. Rescaling their simulation to an initial separation of $100 R_g$, their viscosity was intermediate between our choices of $\nu=0.01\,GM/c$ and $\nu=0.003\,GM/c$ (see Section \ref{sec:numerics}), and their highest-resolution cells have $\Delta r \sim2\,R_g$. \citet{2018MNRAS.476.2249T} used a viscosity roughly equivalent to our $\nu=0.01\,GM/c$ simulation, employed cells as small as $\Delta r\sim0.375 R_g$, and started with an initial separation of $30 R_g$. }  See also recent work along the same vein, which also included the aforementioned post-merger effects \citep{2023arXiv230402575M}.
Additionally, virtually all magnetohydrodynamic simulations of binary black holes, using an approximate but prescribed binary black hole metric or evolving the spacetime dynamically, have chosen initial conditions based on when the binary and disk are expected to decouple \citep[e.g.,][]{2012ApJ...755...51N,2012PhRvL.109v1102F,2014PhRvD..89f4060G,2018PhRvD..97d4036K,2022ApJ...928..187C,2022ApJ...928..137G,2023arXiv230209083R}. 
Overestimating the binary separation at ``decoupling'' by a factor of $\sim2-3$ may cause a mismatch between simulations and the physical systems that they intend to model. This inaccuracy can compound with differences between approximate viscous stresses and those following from magnetohydrodynamic turbulence that may also contribute to binaries decoupling later than predicted \citep[e.g.][]{2012ApJ...755...51N}.

An accurate understanding of when decoupling occurs and the subsequent system evolution is relevant to both present and future observational hunts for binary SMBHs \citep[e.g.][]{2015MNRAS.453.1562G,2016MNRAS.463.2145C,2016MNRAS.461.3145V,2018ApJ...859L..12L,2019ApJ...884...36L,2020MNRAS.499.2245C,2021MNRAS.500.4025L,2021MNRAS.506.2408X}. Furthermore, a more accurate understanding of the decoupling process will be useful for predicting and detecting optical counterparts to inspiraling binaries detected by LISA. Our results suggest electromagnetic counterparts to LISA events that may be sufficient to identify the host galaxies and redshifts thereof, making it possible to constrain the chirp mass of and luminosity distance to each GW source.

\section{Analytical Arguments}\label{sec:analytic}
For simplicity we consider circular binaries composed of objects with masses $m_1$ and $m_2$, for which the rate of change of the semi-major axis of the binary due to gravitational radiation is given in the lowest-order post-Newtonian approximation by \citep{1964PhRv..136.1224P}
\begin{equation}\label{eq:dadtgw}
\frac{da}{dt}=-\frac{64}{5}\frac{G^3m_1m_2(m_1+m_2)}{c^5a_b^3}=\mathcal{A}a^{-3},
\end{equation}
where $\mathcal{A}\equiv64G^3m_1m_2(m_1+m_2)c^{-5}/5$ and $a_b$ is the binary semi-major axis. The evolution of the binary semi-major axis over time from an initial semi-major axis $a_0$ is then 
\begin{equation}\label{eq:atgw}
a_b(t)=a_0\left(1-\frac{4\mathcal{A}}{a_0^4}t\right)^{1/4}= a_0\left(1-\frac{t}{\tau_{\rm GW}}\right)^{1/4}
\end{equation}
where the GW-driven merger timescale is $\tau_{\rm GW}\equiv a_0^4/4\mathcal{A}.$ We denote the total system mass 
$m_1+m_2=M$, and define $R_g=GM/c^2$.  

In a steady state, the rate at which a fluid parcel in a Keplerian accretion disk moves radially inward due to shear stresses, subject to a kinematic viscosity $\nu$ at orbital distance $r$, is given by 
\begin{equation}\label{eq:vrvisc}
v_r = -\frac{3}{2}\frac{\nu}{r}\left(1 + 2\frac{r}{\nu\Sigma}\frac{d(\nu\Sigma)}{dr}\right).
\end{equation}
where $\Sigma$ is the surface density of the disk. In the following estimates we assume globally constant kinematic viscosity and surface density, noting both that some realistic disk models predict only weak dependence of $\nu$ on radius \citep[e.g.][]{2005ApJ...622L..93M} and that the following reasoning holds for arbitrary kinematic viscosity profiles. 
Integrating the position of a fluid element from $r=r_0$ to $r=0$, we find the corresponding timescale to be
\begin{equation}\label{eq:taunu}
\tau_\nu=\frac{1}{3}\frac{r_0^2}{\nu}.
\end{equation}
Many works use an estimate of $\tau_\nu$ which is a factor of $\lambda\sim 4/3$ to $2$ larger, which we will include for the sake of comparison.
\subsection{On Decoupling}\label{sec:decoupling}
One common approximation for when decoupling occurs is found by equating $\tau_{\rm GW}$ with $\tau_\nu(r_c)$, where we parameterize the cavity radius ($r_c$) during the coupled evolution phase as $r_c=\xi a_b$ for some constant $\xi$. The binary semi-major axis at which decoupling occurs is then given by
\begin{equation}\label{eq:adtau}
a_{d,\tau} = \left(\frac{4\lambda\xi^2}{3}\frac{\mathcal{A}}{\nu}\right)^{1/2}.
\end{equation}
However, the semi-major axis evolution described by Equation (\ref{eq:dadtgw}) is relatively slow until $t \approx\tau_{\rm GW}$, given the very sharp dependence of $\dot{a}_b\propto a_b^{-3}$, whereas the viscous spreading of the disk cavity is paced more evenly. For example, starting at a binary separation given by $a_{d,\tau}$, after $\sim0.9\tau_{\rm GW}$, $a_b\approx0.56 a_{d,\tau}$.
On the other hand, the inner edge of the cavity, assuming it spreads only at the rate $v_r=-3\nu/2r$, would have had time to spread to $r_c(t=0.9\tau_\nu)\approx0.32r_c(t=0)$, more than keeping pace with the binary. 
Equation (\ref{eq:adtau}) will thus overpredict the separation at which decoupling occurs unless the disk takes longer than $\tau_\nu$ to respond to the binary inspiral.

A more appropriate criterion for decoupling is reached by equating $\dot{a}_b$ with $v_r(r_c)$ \citep[e.g.][]{2002ApJ...567L...9A}, resulting in the following approximation for the semi-major axis at which disk and binary decouple:
\begin{equation}\label{eq:advel}
a_{d,\dot{a}}=\left(\xi\frac{2}{3}\frac{\mathcal{A}}{\nu}\right)^{1/2}=\frac{a_{d,\tau}}{\sqrt{2\lambda\xi}}.
\end{equation}
We demonstrate in Section \ref{sec:results} that Equation (\ref{eq:advel}) is significantly more accurate than Equation (\ref{eq:adtau}) when compared to our hydrodynamical simulations. Although a factor of $\sqrt{2\lambda\xi}$ is certainly within an order of magnitude in accuracy, taking $\xi\sim2-3$ (as seen in simulations) and (conservatively) $\lambda=1$, this results in a factor of $\sim2-3$ error in $a_d$ and a factor of $\sim3-4$ error in the estimated orbital period at which the binary decouples.

\subsection{On the Presence of Gas After Decoupling}\label{sec:gaspres}
While the binary and disk are coupled, each SMBH will have within its Roche lobe an accretion disk of its own (hereafter a `minidisk'). Considering an equal-mass binary, the radius of each minidisk can be approximated as $\sim\delta a_b$ where $\delta\approx0.379$ \citep[e.g.][]{1983ApJ...268..368E}. When $a_b=a_{d,\dot{a}}$ the viscous timescale at the edge of the minidisks is $\tau_m\sim 2 \delta^2 \xi \mathcal{A}/(9\nu^2)$ and the GW inspiral timescale is $\tau_{\rm GW}\sim \xi^2 \mathcal{A} / (9 \nu^2)$.  Thus, as long as $\tau_m / \tau_{\rm GW} \sim 2 \delta^2 / \xi < 1$, the timescale for the minidisks to viscously accrete after decoupling (at $a_{d,\dot{a}}$) will be shorter than the timescale for the binary to coalesce.

The above argument suggests that for gas to be present during merger, a binary must continue accreting from its circumbinary disk after the nominal point of decoupling given by Equation (\ref{eq:advel}). Binaries are able to strip matter from the cavity walls on a dynamical timescale, so gas may continue accreting if decoupling occurs proximate to the merger. However, if the binary decouples from the disk well before merging, the accretion rate should eventually plummet as the binary and cavity lose contact with one another.

\begin{figure*}
\includegraphics[width=\linewidth]{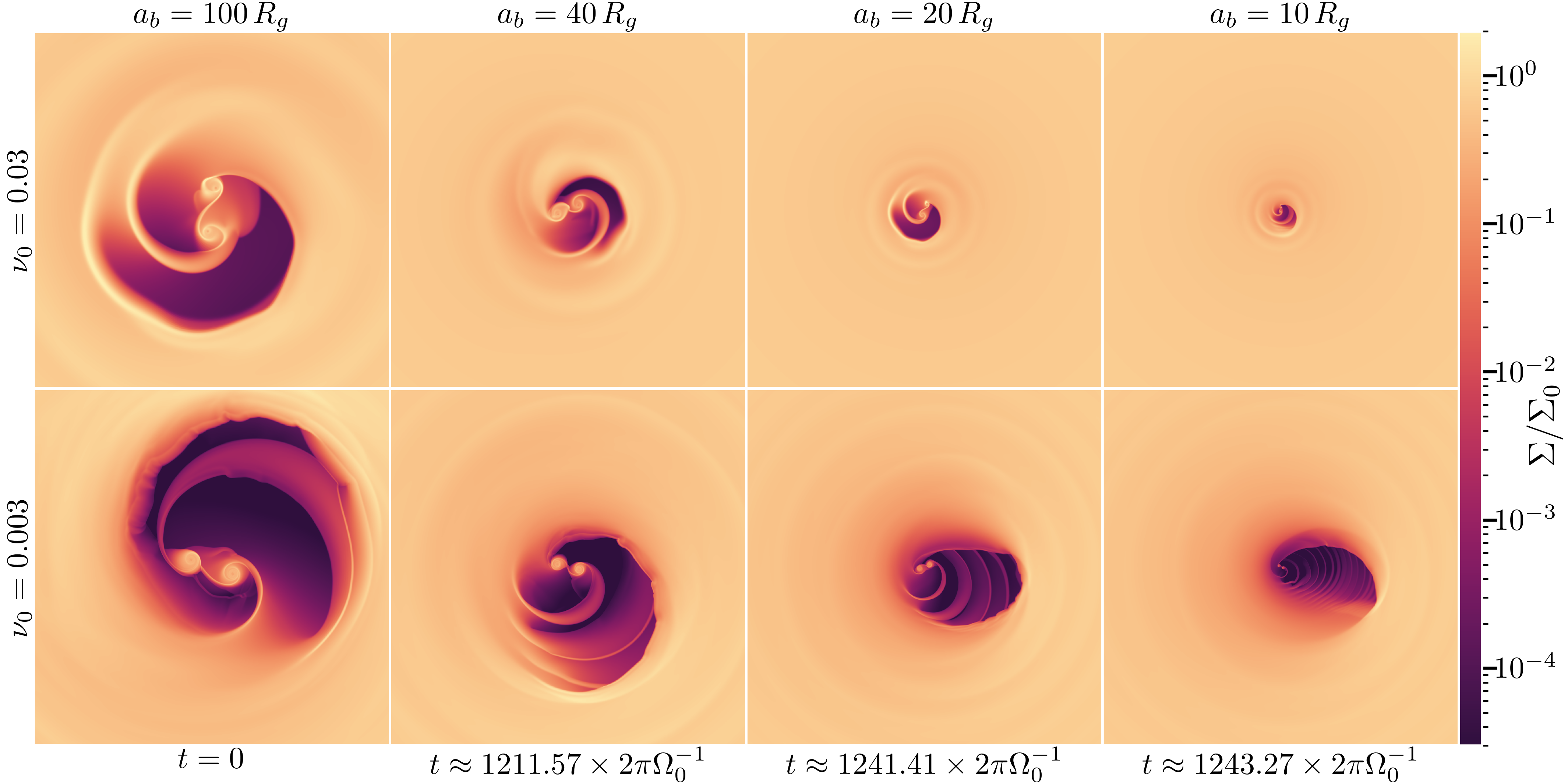}
\caption{Surface density profiles from our $\nu_0=0.03$ and $\nu_0=0.003$ simulations over the course of their inspirals, during which decoupling occurs around $a_b\sim14\,R_g$ and $a_b\sim56\,R_g$ respectively. Each panel is $800\,R_g$ long in each dimension.  
The flow of gas around the binary near merger is qualitatively different when the system decouples at earlier times.
}
\label{fig:surface}
\end{figure*}

\section{Numerical Methods}\label{sec:numerics}

To quantitatively explore circumbinary disk decoupling, we carried out a suite of simulations equal mass binaries using the \texttt{Athena++}\citep{2020ApJS..249....4S} code to solve the vertically-integrated Navier-Stokes equations in Cartesian coordinates. See \citet{DR2023} for a detailed explanation of our numerical methods. We initialized each binary with a separation of $a_0=100 R_g$, and simulated disks with $\nu=\nu_0 GM/c$, where $\nu_0\in\{0.1, 0.03, 0.01, 0.003, 0.001\}$.\footnote{In terms of initial binary orbital parameters, $GM/c=a_{0}^2\Omega_0/10$, and thus our lowest viscosity was $\nu=10^{-4}\,{\rm a_{0}^2\Omega_0}$. In comparison to $\alpha-$viscosity models \citep{1973A&A....24..337S}, which posit that $\nu=\alpha c_sH$ where $H$ is the disk scale height, or $\nu=\alpha(H/r)^2\sqrt{GMr}$ for a disk with constant $H/r$, $\nu_0\sim10\alpha(H/r)^2$ at $r=a_0=100R_g$. Concretely, for disks with $H/r\sim1/10,$ our $\nu_0=0.01$ model is roughly comparable to one with $\alpha=0.1$.}
In this parameterization, Equation (\ref{eq:advel}) reduces to 
\begin{equation}\label{eq:decoupleM}
a_d=\sqrt{\frac{32\xi}{15\nu_0}}R_g
\end{equation}

We prescribed the motion of the binary analytically, using Equation (\ref{eq:atgw}) and the equations for the evolution of the binary angular velocity and orbital phase that follow therefrom. We use a softened gravitational potential for each black hole of $\Phi_i = -Gm(r_i^2+\epsilon_{g}^2)^{-1/2},$ where $\epsilon_g={\rm max}(0.033a_b,r_g)$ is a gravitational softening length and $r_i$ is the distance between a fluid element and black hole $i$. We introduce a sink term for each black hole which removes gas at a rate $600(0.05a_0/r_s)^2\nu_0\Omega_0(1-(r_i/r_s)^4)^4$ if $r_i<r_s$ and $0$ otherwise, where the sink radius $r_s={\rm max}(0.05a_b,6r_g)$, $r_g=GM/2c^2$, and $\Omega_0=\sqrt{GM/a_0}$, so that at $r_s$ the mass removal timescale is a few times shorter than $\tau_\nu$.
We use torque-controlled sinks \citep{2021ApJ...921...71D}, removing specific spin 
angular momentum from accreting gas, with respect to each SMBH, equal to that at the innermost stable circular orbit ($6r_g$) 
if $r_i>6r_g$ and not altering fluid velocities if $r_i\leq6r_g$. 
We implemented damping zones at the outer boundary and a locally isothermal equation of state similarly to \citet{2019ApJ...875...66M}, setting the azimuthal Mach number to $\mathcal{M}=10$. We used the Roe approximate Riemann solver \citep{1981JCoPh..43..357R}, third-order reconstruction \citep{2018JCoPh.375.1365F}, and the second-order van Leer (VL2) time integrator \citep{2009NewA...14..139S}.
We utilize the adaptive mesh refinement infrastructure within \texttt{Athena++} to increase resolution as the binary contracts, maintaining a cell aspect ratio of $\Delta x/r=\Delta y/r\approx (2/300){\rm{max}}(a_b, r)$ when possible as long as $r>{\rm max}(a_b, 20r_g).$ 

We initialized each disk with a surface density profile of $\Sigma=\Sigma_0\exp[-(r/2.5a_0)^{-12}][1-0.7\sqrt(a_0/r)]$ \citep[see, e.g.][]{1974MNRAS.168..603L,2017MNRAS.466.1170M,2022MNRAS.513.6158D},
a radial velocity of $-3\nu/2r$, and an angular velocity profile of 
\begin{equation}
\Omega^2(r)=\frac{GM}{a_0^3}\left(1+\frac{3}{16}\frac{a_0^2}{r^2}\right)+\frac{1}{r\Sigma}\frac{d\Pi}{dr},
\end{equation}
where $\Pi$ is the vertically-integrated pressure, in approximate hydrostatic equilibrium in the Newtonian quadrupole approximation of the binary potential. The accretion rate through the disk is $\dot{M}=(3\pi\nu\Sigma$, which even for super-Eddington accretion rates corresponds to a total disk mass within our computational domain $(r\lesssim1000\,R_g)$ many orders of magnitude below the mass of the binary. Accordingly, we do not include the effects of disk self-gravity, and our simulations are scale-free.
We allowed each simulation to settle into a quasi-steady state before allowing the binary to inspiral. These burn-in periods lasted ${\rm min}[300, 3/\nu_0]2\pi\Omega_0$.

\section{Results}\label{sec:results}
We first demonstrate how the morphology of the binary-disk system changes over the course of decoupling for different disk viscosities. Subsequently, we illustrate related changes in variability, and the potential observable signatures of decoupling. 

For the sake of brevity we do not report here a detailed analysis of gas torques on the binary during the inspiral. However,  over the course of decoupling gravitational torques between the binary, circumbinary disk, and streams of gas within the cavity, which are largely responsible for negative contributions to the gravitational torque on the binary from the gas \citep[e.g.,][]{2020ApJ...900...43T,2022MNRAS.513.6158D}, become weaker as the binary decouples. The minidisks and their positive torque contributions remain, so the gravitational torque on the binary tends to increase by a factor of order unity (normalized by the accretion rate and binary angular momentum $\dot{M}J_b/M$) during the decoupling process, although this effect may be more substantial in thinner disks. 

\subsection{Disk Morphology}\label{sec:morph}
Surface density profiles from our $\nu_0=0.03$ and $\nu_0=0.003$ simulations are shown in Figure \ref{fig:surface} at a few times after the initial burn-in. 
As expected, the cavity of the higher-viscosity circumbinary disk is generally smaller, and tracks the evolution of the binary until later in the inspiral. 
Although whether or not the binary and disk have decoupled by $a_b=10\,R_g$ in the $\nu_0=0.03$ simulation is not visually obvious, the evolution of the circumbinary disk has clearly decoupled from the binary by $a_b \approx 20\,R_g$ in the $\nu_0=0.003$ case, although some gas is still present in the cavity.
Additionally, even though the binary has clearly decoupled from the circumbinary disk by $a_b=10\,R_g$ in the $\nu_0=0.003$ simulation, the minidisks are still present, although their surface densities have decreased over time.

\begin{figure}
\includegraphics[width=\linewidth]{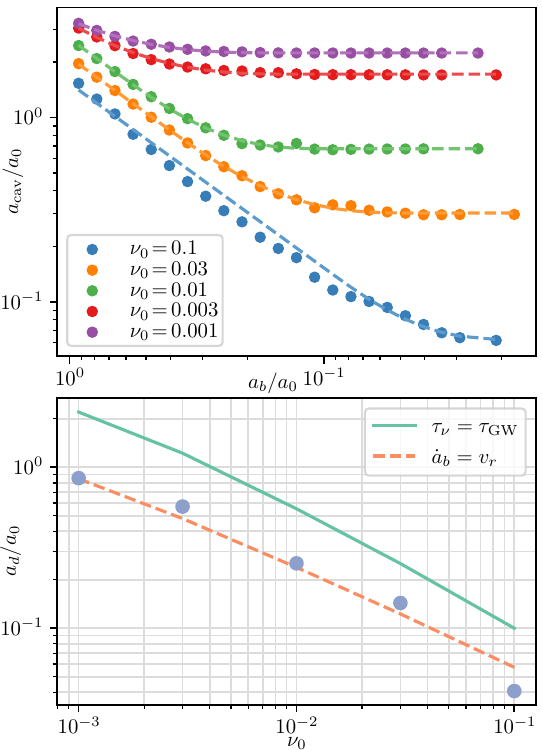}
\caption{
The top panel displays the measurements of the cavity semi-major axis as a function of binary semi-major axis from each of our simulations. The bottom panel displays our inferences of the characteristic binary semi-major axis at decoupling compared with timescale-based predictions (Equation (\ref{eq:adtau}), a solid green line) and velocity-based predictions (Equation (\ref{eq:advel}), a dashed orange line). Although decoupling is a gradual process, the characteristic binary semi-major axis at which it occurs is described well by Equation (\ref{eq:advel}).}
\label{fig:cavity}
\end{figure}

\begin{figure*}
\includegraphics[width=\linewidth]{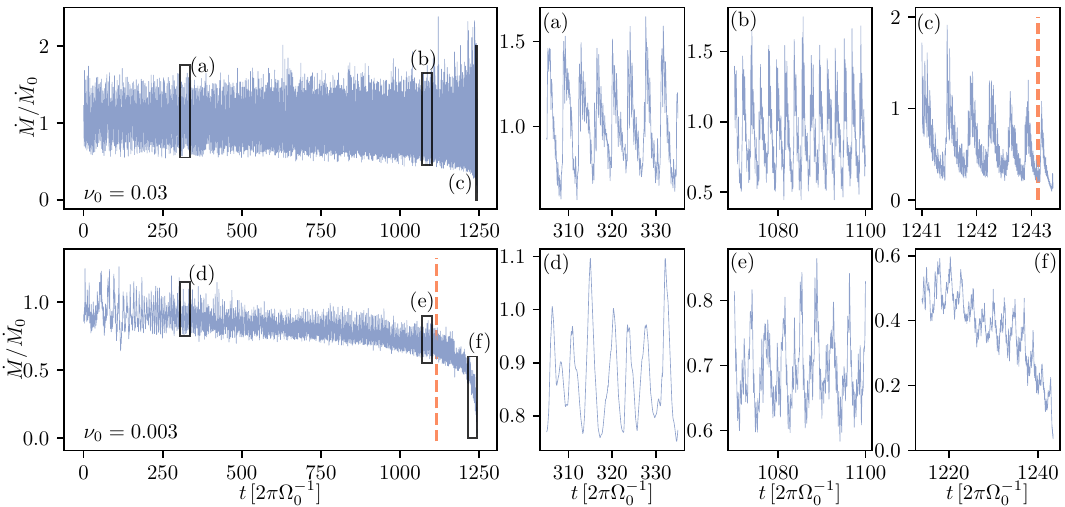}
\caption{The accretion rates, normalized by $\dot{M}_0=3\pi\nu\Sigma_0$, for our $\nu_0=0.03$ and $\nu_0=0.003$ simulations. The leftmost panels display the accretion rates over the course of the entire simulations, after the initial burn-in stage. The other panels in each row focus on narrower windows, displaying shorter-term variation in the binary accretion rate. Vertical dashed lines indicate the time at which each binary reaches $a_b=a_d$. We note that because the initial binary orbital period $\Omega_0=10^{-3}c^3G^{-1}M^{-1}$, $2\pi\Omega_0^{-1}\approx8.6(M/10^6\,M_\odot)$ hours. While the accretion rate in the high-viscosity case remains significant until very late in the inspiral, sharply declining accretion rates in systems which decouple at larger separations should identify such systems.}
\label{fig:accretion}
\end{figure*}

We measured the evolution of the circumbinary disk cavity using time-averaged histograms of the fluid surface density, binned according to fluid element semi-major axis \citep[see, e.g. Figure 17 of][]{2022MNRAS.513.6158D}. We defined the semi-major axis of the cavity ($a_{\rm cav}$) as that at which surface density reaches a value of $\Sigma_0/5$ at $a>a_b$. 
Although our definition of cavity semi-major axis is somewhat arbitrary, we have verified that our results are insensitive to the precise definition used. The top panel of Figure \ref{fig:cavity} presents the results of these measurements for each viscosity, along with nonlinear least-squares fits of a smoothly broken power law shown using dashed lines. For our broken power law fit, we fix early- and late-time power law indices $a_{\rm cav}\propto a_b^p$ to $p=1$ and $p=0$ respectively, fitting for a break location, transition width, and constant of proportionality. We define the binary semi-major axis which characterizes the decoupling of the binary from the circumbinary disk ($a_d$) as the location of the power-law break. Although $a_d$ is a characteristic point in the evolution of the binary, Figure \ref{fig:cavity} also illustrates that decoupling is a gradual process as opposed to an instantaneous one. 

The bottom panel of Figure \ref{fig:cavity} compares 
our measurements with the predictions made by Equations (\ref{eq:adtau}) and (\ref{eq:advel}) using $\lambda=1$ and measuring $\xi$ at the end of the burn-in stage of each simulation.
As expected based on Section \ref{sec:decoupling}, timescale-based reasoning  overestimates the characteristic binary semi-major axis at which decoupling occurs. Additionally, Equation (\ref{eq:advel}) makes fewer and weaker assumptions than Equation (\ref{eq:adtau}), not depending on $\lambda$ and depending more weakly on $\xi$. Comparing the values of $a_d$ in Figure \ref{fig:cavity} to the surface density snapshots in Figure \ref{fig:surface} illustrates that the disk-binary system continues to evolve together after this nominal decoupling point ($a_d\approx56\,R_g$ for the $\nu_0=0.003$ simulation), albeit at a slower rate.

\subsection{Accretion}\label{sec:accretion}
Inspiraling binaries will likely emit radiation as long as gas is present around either SMBH. Here we investigate the degree to which the magnitude and periodicity of accretion onto the binary vary throughout the decoupling process. 
Although our simulations cannot probe emission directly, studies have found that the accretion rate onto binaries can correlate with their luminosity \citep[e.g.][]{2015MNRAS.446L..36F} and Poynting flux \citep{2022ApJ...928..187C}.
Figure \ref{fig:accretion} illustrates how the accretion rate changes over the course of the inspirals in our $\nu_0=0.03$ and $\nu_0=0.003$ simulations, which decouple at separations of $a_d\approx14\,R_g$ and $a_d\approx56\,R_g$ respectively. The times at which each binary reaches $a_d$ are indicated by vertical dashed lines. In both cases, the dominant period on which accretion varies is typically is the orbital period of the cavity. 
\begin{figure*}
\includegraphics[width=\linewidth]{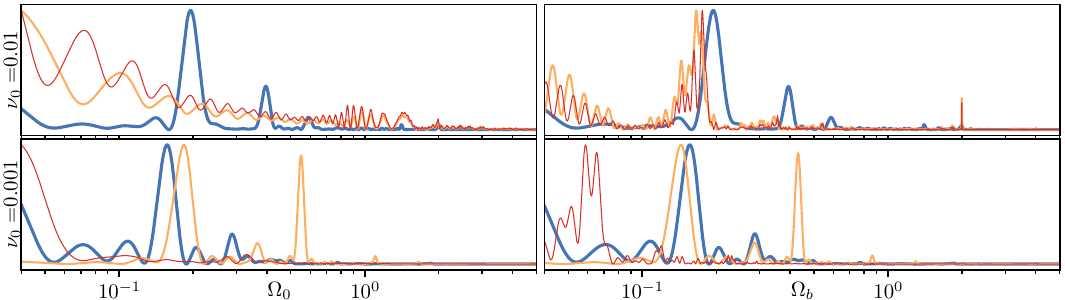}
\caption{Lomb-Scargle periodograms of the accretion rate time series from our $\nu_0=0.01$ and $\nu_0=0.001$ simulations during three epochs: blue lines plot the periodograms over the first $50\pi\Omega_0^{-1}$ after burn-in, orange lines plot the periodograms over a period of $50\pi\Omega_0^{-1}$ centered on the time of decoupling, and red lines plot the periodogram over the final $50\pi\Omega_0^{-1}$ of each simulation before merger. The left column plots periodograms in the lab frame of the binary. Because the binary orbital frequency changes rapidly at late times, we also plot in the right column periodograms of a rescaled accretion rate time series, in terms of the instantaneous binary frequency and orbital phase. All periodogram amplitudes are renormalized to peak at 1. Accretion rate periodicity continues until late in the inspiral in systems which decouple at later times, whereas in lower-viscosity disks virtually no significant periodicity persists.}
\label{fig:lsp}
\end{figure*}
Comparing panels $(a)$ and $(b)$ of Figure \ref{fig:accretion}, the average accretion rate stays roughly constant but the time series varies at a considerably higher frequency. Comparing panels $(d)$ and $(e)$, which highlight the same points in the binary inspiral as $(a)$ and $(b)$ but for the lower-viscosity disk, the mean accretion rate decreases slightly; and the accretion rate exhibits higher-frequency variability, but not to the same extent as in the higher-viscosity system. As the $\nu_0=0.03$ case decouples quite late in the inspiral, plenty of gas is present throughout the merger and the accretion flow is predominantly a scaled-down version of that at larger separations. In the lower-viscosity case however, the accretion rate drops precipitously after decoupling. Although the binary is still able to accrete for some time after reaching $a_d$, as it continues to leave the disk behind the accretion rate eventually drops by a factor of $\gtrsim 5$ over a period of $\sim10(M/10^6M_\odot)\,\rm{days}$. 
If such a significant decrease in accretion rate is accompanied by a commensurate decrease in flux, particularly at shorter 
wavelengths, it may be possible to uniquely identify the host galaxy of an event within the error volume of an ongoing LISA event.

We further investigate periodicity in systems throughout decoupling by constructing Lomb-Scargle periodograms \citep{{1976Ap&SS..39..447L},{1982ApJ...263..835S},{2010ApJS..191..247T}} of the accretion rate times series during different epochs for our $\nu_0=0.01$ and $\nu_0=0.001$ simulations: the former straddles the boundary in behavior, decoupling around $a_d\sim25\,R_g$ and still exhibiting significant long-term variability on a variety of timescales throughout the merger despite a declining accretion rate; the latter decouples around $a_d\sim85\,R_g$, and generally exhibits little discernible periodicity at late times. These periodograms are shown in Figure \ref{fig:lsp}, after rescaling each time series to have zero mean and unit variance, and renormalizing the resulting power spectra to have unit amplitude. In each panel of Figure \ref{fig:lsp}, different colors illustrate periodicity over different windows: the first and final $50\pi\,\Omega_0^{-1}$ of each simulation are shown in blue and red respectively, and the periodogram over the window of $50\pi\,\Omega_0^{-1}$ centered on the time of decoupling is shown in orange. The left columns plot periodicity in terms of the initial binary orbital period, and the right columns plot periodicity in terms of the current binary orbital period, rescaling each time series according to the binary orbital phase.

Considering the $\nu_0=0.01$ simulation, periodicity at early times is primarily at $\sim0.2\Omega_0$, roughly the orbital frequency of the cavity wall, although there is some power at its higher harmonics and twice the binary angular frequency. Since the $\nu_0=0.01$ simulation decouples $\sim30\Omega_0^{-1}$ before $\tau_{\rm GW}$, the data near decoupling and prior to merger overlap substantially. In both cases variability associated by the cavity is largely washed out in lab-frame frequency by the secular decrease in the accretion rate throughout decoupling.\footnote{Because $\Omega_b\geq\Omega_0$, the decrease in accretion rate occurs over a much larger change in binary orbital phase than $2\pi\Omega_0^{-1}$, decreasing the low-frequency power at late times when measured in terms of the instantaneous binary orbital frequency. Similar trends have been observed in \citep[e.g.][]{2018MNRAS.476.2249T,2019ApJ...879...76B}.} In terms of the binary orbital frequency however, the power spectrum peak associated with the cavity orbital period moves to lower frequencies as the binary decouples and the orbital frequency of the binary increases. In the $\nu_0=0.01$ simulation we always observe strong periodicity on the binary orbital period. 

In the $\nu_0=0.001$ case, periodicity is dominated by the orbital period of the disk cavity and its harmonics as has been observed previously \citep[e.g.][]{2014ApJ...783..134F,2021ApJ...922..175N}. Although in the lab frame the cavity orbital frequency slightly increases between the beginning of the simulation and decoupling, as the cavity shrinks, the frequency relative to the binary decreases, and the binary evolution outpaces that of the disk. At late times, almost all lab-frame variability is dwarfed by the secular decline in the accretion rate. Our averaging window was also large enough, in terms of the binary orbital period, to identify some low-frequency periodicity due to accretion from the cavity walls. However, even in terms of the evolving binary period, there is virtually no variability on the orbital period of the binary. In these cases, however, effects such as Doppler modulation \citep[e.g.,][]{2017MNRAS.470.1198D} and self-lensing \citep[e.g.,][]{2022PhRvD.105j3010D} might lead to variability in observed light curves. 

\subsection{Gravitational Waves}
Inspiraling SMBH binaries are expected to be one of the most prominent signals detectable by LISA. As illustrated above, gas may be present around the binaries throughout their inspiral, and it therefore may be possible to identify electromagnetic counterparts corresponding to LISA signals. However, the types of variability corresponding to each gas-rich merger depend on when each binary-disk system decouples. Additionally, the torques resulting from interactions between the binary and the gas, while necessarily subdominant in the decoupling regime, may leave subtle imprints on the gravitational waveform. 

\begin{figure}
\includegraphics[width=\linewidth]{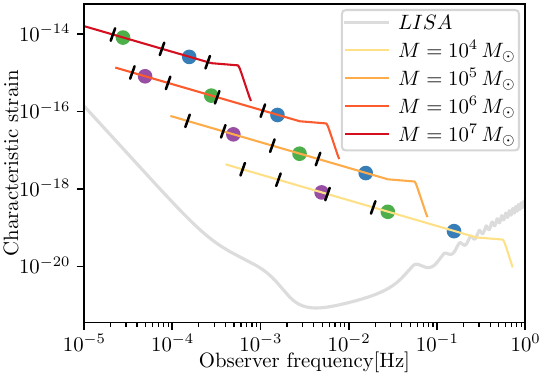}
\caption{The evolution of black hole binaries in the LISA band at $z=2$ over the final three years prior to merger. Purple, green, and blue dots mark the points where each binary reaches $a_d$, given by Equation \ref{eq:decoupleM}), for $\nu_0=\{0.001,0.01, 0.1\}$ respectively. The dashed markings along each line mark one year, one month, one day, and one hour before merger. Depending on the system mass and disk viscosity, many binaries might decouple from their disks while emitting in the LISA band. This could identify the host galaxy of the merger, facilitating multi-messenger observations.}
\label{fig:lisa}
\end{figure}
We illustrate where in the LISA band decoupling should occur for a few black hole binary systems at $z=2$ in Figure \ref{fig:lisa}. Along with an approximate LISA sensitivity curve \citep{neil_cornish_2019_2636514,2019CQGra..36j5011R} and phenomenological inspiral-merger-ringdown models \citep[PhenomA,][]{Ajith_2007}, we mark the characteristic points of decoupling for each inspiral for $\nu_0=\{0.1,0.01,0.001\}$ according to Equation (\ref{eq:decoupleM}). 
For a given $\nu_0$ and $\xi$, $a_d/R_g$ is constant, causing the characteristic GW and orbital frequencies at decoupling to be inversely proportional to the binary mass.
Similarly, this causes the the time between decoupling and merger to be proportional to the binary mass, requiring higher-cadence observations to observe decoupling in lower-mass black holes.
In principle, assuming that the GW signal can reliably measure the mass of a binary, observational signatures of decoupling may be able to constrain the effective viscosity of circumbinary accretion disks. 

To estimate the effect of gas on the phase of GWs \citep[e.g.][]{2021MNRAS.501.3540D,2022MNRAS.517.1339G,2023MNRAS.tmp..685Z}, we parameterize the rate of binary orbital evolution as $\dot{a}_b\approx\aleph a_b\dot{M}/M$, where we measure $\aleph\sim2-3$ in our present simulations, but $\aleph$ decreases sharply with Mach number in the coupled regime \citep[e.g. $\aleph\sim-90$ at $\mathcal{M}=50,$][Section 4.3]{2022MNRAS.513.6158D}, although during decoupling $\aleph$ should increase as streams of gas in the cavity become dynamically negligible. We parameterize the accretion rate as some fraction of the Eddington-limited rate $\dot{M}=\eta\dot{M}_{\rm Edd}$, where $\dot{M}_{\rm Edd}\equiv4\pi GMc^{-1}\epsilon^{-1}\kappa^{-1}$ where we will assume a Thomson opacity $\kappa=0.4\,{\rm cm^2\,g^{-1}}$ and the efficiency at which rest mass is converted to radiation $\epsilon$ to be 10\%. The dynamical importance of gas on the inspiral can be characterized using the ratio 
\begin{equation}
\frac{\dot{a}_{\rm gas}}{\dot{a}_{\rm GW}} = -\frac{20\pi}{16}\frac{\eta\aleph}{\epsilon}\frac{1}{\kappa}\frac{c^4a^4_b}{M^3G^2}=-\frac{20\pi}{16}\frac{\eta}{\epsilon}\frac{\aleph}{\kappa}\frac{n^4G^2M}{c^4},
\end{equation}
where on the right-hand side we have defined $n\equiv a_b/R_g$. This ratio is small ($\lesssim10^{-2}$) when $n\lesssim3\times10^3|\aleph|^{-1/4}(M/10^6M_\odot)^{-1/4}$, which is satisfied well before decoupling for most viscosities.

The accumulated phase of a GW event for a circular binary is $\phi=2\pi\int_{a_i}^{a_f}f(a_b)da_b/\dot{a}_b$ between initial and final separations $a_i$ and $a_f$. When $\dot{a}_{\rm gas}\ll\dot{a}_{\rm GW}$ the cumulative deviation in phase due to gas interactions is approximately $\delta\phi\approx2\pi\int_{a_i}^{a_f}f(a_b)\dot{a}_{\rm gas}/\dot{a}^2_{\rm GW}$. Then, because 
\begin{equation}
\frac{f}{\dot{a}_{\rm GW}}=\frac{5}{16\pi}\frac{c^5a^{3/2}}{M^{5/2}G^{5/2}} 
\end{equation}
the cumulative phase shift (relative to a merger in vaccum) is
\begin{equation}
\delta\phi\approx-\frac{25\pi}{208}\frac{\eta\aleph}{\epsilon\kappa}\frac{c^9(a_f^{13/2}-a_i^{13/2})}{M^{11/2}G^{9/2}}.
\end{equation}
For a GW-driven inspiral over an observing window $\Delta t$, $a_f=a_i(1-\Delta t/\tau_{\rm GW})^{1/4}$, so for an initial separation of $a_i=NGM/c^2$. 
\begin{equation}
\delta\phi = \frac{25\pi}{208}\frac{\eta\aleph}{\epsilon\kappa}\frac{G^2M}{c^4}N^{13/2}\left[1-\left(1-\Delta t/\tau_{\rm GW}\right)^{13/8}\right].
\end{equation}
The maximum phase shift will occur if a full inspiral is observed, in which case, from Equation (\ref{eq:atgw}), $N\approx125(\Delta t/3\,{\rm yr})^{1/4}(M/10^6\,M_\odot)^{-1/4}$, resulting in 
\begin{equation}
\delta\phi\approx4\times10^{-4}\aleph\left(\frac{\eta}{0.1}\right)\left(\frac{\Delta t}{3\,{\rm yr}}\right)^{13/8}\left(\frac{M}{10^6\,M_\odot}\right)^{-5/8}.
\end{equation}

A phase modification should be detected reliably if $\delta\phi\geq 10/\rho$ \citep{2011PhRvD..84b4032K}, where $\rho$ is signal-to-noise ratio of the GW event. Inspiraling binaries at $z\lesssim4$ with chirp masses between $\sim10^{4.6}\,M_\odot$ and $\sim10^{6.8} \,M_\odot$ should be detected with $\rho\gtrsim300$, or $\rho\gtrsim3000$ for binaries with source-frame chirp masses near $\sim10^6\,M_\odot$ and redshifts below $z\lesssim2$ \citep{2019MNRAS.486.4044B}. 
Thus, LISA may be able to reliably detect gaseous modifications to the gravitational radiation of SMBH binaries. 

\section{Summary}\label{sec:summary}
We have investigated the process by which binaries inspiraling due to GW emission decouple from their circumbinary disks. We showed analytically and via simulations (see Figure \ref{fig:cavity}) that while this process is gradual, as opposed to sudden, it occurs at a characteristic binary semi-major axis given by Equation (\ref{eq:advel}), when the viscous inflow rate of the disk matches the rate of change of the binary semi-major axis -- a factor of $\sim3$ smaller than estimates based on equating $\tau_{\rm GW}$ and $\tau_\nu$, corresponding to differences in the orbital period at decoupling of $\sim3-4$.

Higher-viscosity $\nu\gtrsim0.03\,GM/c$ disks decouple very near merger ($a_b\lesssim15\,R_g$) and exhibit variability and accretion morphology similar to coupled binaries well into the highly-relativiscic regime. Lower viscosities $\nu\lesssim0.01\,GM/c$, appropriate to thin ($\mathcal{M}\gtrsim10$) disks \citep{1973A&A....24..337S}, lead to decoupling at wider separations. In such systems the accretion rate sharply declines approaching merger and the binary leaves behind a large eccentric cavity, accreting primarily from its minidisks with negligible periodicity. We showed where in the LISA band various binaries decouple, and illustrated that while dynamically inconsequential, gas can leave a measureable imprint on the GW phase observed by LISA. 

\section*{Software}

\texttt{Athena++} \citep{2020ApJS..249....4S}, \texttt{matplotlib} \citep{4160265}, \texttt{cmocean} \citep{cmocean}, \texttt{numpy} \citep{5725236}, \texttt{yt} \citep{2011ApJS..192....9T}

\section*{Acknowledgments}
We thank the referee for their constructive report, which helped clarify the presentation of our results. The authors acknowledge the University of Maryland supercomputing resources (http://hpcc.umd.edu) that were made available for conducting the research reported in this paper, and the ASTRA computing cluster
maintained by the Department of Astronomy at the University of Maryland.
AJD and MCM are supported in part by NASA ADAP grant 80NSSC21K0649. Research at Perimeter Institute is supported in part by the Government of Canada through the Department of Innovation, Science and Economic Development and by the Province of Ontario through the Ministry of Colleges and Universities.

\bibliographystyle{aasjournal}
\bibliography{references}
\end{document}